\documentclass[12pt]{article}

=-0.5cm \mathsurround =1pt \hoffset =0.5cm \relpenalty=10000
\def\be{\begin{equation}}
\def\ee{\end{equation}}
\def\ed{\end{document}}
\def\ba{\begin{eqnarray}}
\def\ea{\end{eqnarray}}

\def\r{\mathbf r}
\def\v{\mathbf v}
\def\p{\mathbf p}

\def\ko{\mbox{\it\tiny{Ko}}}
\def\ze{\mbox{\it\tiny{ZeLe}}}

\begin{document}


\author{\bf G. Erochenkova\and\bf C. Chandre}

\title{\bf PHOTON PLASMA-WAVE INTERACTION VIA COMPTON SCATTERING}

\date{}
\maketitle
\font\fifteen=cmbx10 at 15pt \font\twelve=cmbx10 at 12pt
\begin{center}
{\bf Centre de Physique Th\'{e}orique - UMR 7332, Luminy  Case 907, \\
13288 Marseille, Cedex09, France}

\vspace{0.5 cm}
{\bf {Abstract}}
\end{center}
\vspace{0.5cm}
{\small The Kompaneets theory of photon kinetic evolution due to the  Compton effect is extended to the case of the
Vlasov plasma wave oscillations. \\Taking into account Zel'dovich-Levich's approximation we study interaction of accumulating photons with plasma  in the long wavelength limit.}

\vspace{2cm}

{\bf Introduction}

\vspace*{0.5cm}
\noindent To consider the role of the Compton scattering of quanta on non-relativistic electrons, Kompaneets proposed~\cite{KO}
the following kinetic equation for the photon-number distribution
function $n_{\ko}(\hbar\omega,t)$ in a unit volume:
\begin{eqnarray}\label{a1}
&&\frac{\partial n_{\ko}}{\partial t}=-\int {d^3 p}\int
dW ~ n_{\ko}(\hbar\omega,t)(1+n_{\ko}(\hbar\omega^\prime,t))f_{0}(\varepsilon)\\
&&+\int {d^3p}\int dW  ~ n_{\ko}(\hbar\omega^\prime,t)(1+n_{\ko}(\hbar\omega,t))f_{0}(\varepsilon+\hbar(\omega-\omega^\prime));\nonumber
\end{eqnarray}
where $n_{\ko}(\hbar\omega,t)=
\int {d^3r}~ n_{\ko}(\hbar\omega,{\r},t)$,
and the integrand is the photon density for a given energy $\hbar\omega$ and time $t$.
Here $f_{0}(\varepsilon)$ is the Maxwell distribution function for temperature $T_e$ of free electrons, $\varepsilon=p^2/(2m)$; $dW$  is the differential photon transition probability  from one state into another due to scattering  with electrons. In this equation, emission and absorption processes have not been taken into account, so that the transitions are produced exclusively by Compton scattering processes.
It is easy to check that in stationary case $\partial_t\overline{n}_{\ko}=0$ the solution is the Planck distribution:
$\overline{n}_{\ko}(\hbar\omega)= 1/(\exp(\hbar\omega/kT)-1)$.\\
Kompaneets considered the non relativistic case, i.e., it was assumed that the inequality $ kT \ll mc^2$ holds.
Introducing a new variable $\hbar\omega/kT=x$ and notation $\omega^\prime-\omega=\Delta$ he got from~(\ref{a1}) in approximation $|\triangle|<<\omega$:
\begin{eqnarray}\label{a11}
&&\frac{\partial n_{\ko}}{\partial t}\!\! =\!\! \left(\frac{\partial n_{\ko}}{\partial x}+n_{\ko}(1+n_{\ko})\right)
 \frac{\hbar}{kT}\int\! {d^3 p}\int\! dW f_0(\varepsilon)\Delta\!\!\!\!\nonumber\\
&&\!\!+\!\! \left(\frac{{\partial}^2 n_{\ko}}{\partial x^2}\!+\!2(1+n_{\ko})\frac{\partial n_{\ko}}{\partial x}\!+\!n_{\ko}(1\!+\!n_{\ko})\right )
  \frac{1}{2}\!\frac{\hbar^2}{(kT)^2}\!\int\! {d^3 p}\int\! dW f_0(\varepsilon){\Delta}^2\!.
 \end{eqnarray}
After introduction of  dimensionless time parameter $t_1=(mc^2/kT)(l/c)t$, where $l$ is the Compton range determined by the total cross section $ 8\pi e^2/(3mc^2)$, and calculation of integrals in~(\ref{a11}), Kompaneets obtained the following equation:
\begin{equation}\label{kz1}
  \frac{\partial n_{\ko}}{\partial t_1}=\frac{1}{x^2}\frac{\partial}{\partial x}\left[  x^4
  {\left(\frac{\partial n_{\ko}}{\partial x}+n_{\ko}+n_{\ko}^2\right)}\right].
\end{equation}
Later Zel'dovich and Levich~\cite{ZL} found solution of Eq.~(\ref{kz1}) in the limit $n_{\ko}\gg 1$ (high-temperature regime)  and $n_{\ko}^2\gg |\partial n_{\ko}/\partial x|$, since in this regime Eq.~(\ref{kz1}) reduces to the inviscid Burgers' equation:
\begin{equation}\label{z1}
\frac{\partial n_{\ze}}{\partial t_1}=\frac{1}{x^2}\frac{\partial}{\partial x}( x^4n_{\ze}^2).
\end{equation}
They showed that in the absence of absorption the photons undergo a kind of  Bose condensation in the energy space in the vicinity of zero. This  kinetic condensation  depends essentially on the form of the initial photon distribution. For a certain form of initial distribution,
 a shock-wave as a function of photon energy occurs in the course
 of its dynamics. The process is extremely non-uniform across the frequency spectrum and substantially affected by absorption.
 Using the method of characteristics Zel'dovich and Levich have found solution of Eq.~(\ref{z1}) in the following form:
\begin{equation}\label{z2}
 x=F(x^2n_{\ze})-2t_1x^2n_{\ze}.
\end{equation}
Here $F$ is determined by the initial condition for Eq.~(\ref{z1}).
According to Eqs.~{(\ref{z1})-(\ref{z2})} all points on the initial curve $F(x)=x^2 n_{\ze}(x,t=0)$ move along characteristic straight lines parallel
  to the $x$-axis in the direction of decreasing $x$ with velocity proportional to $x^2n_{\ze}$. The time at which a given point reaches dimensionless energy $x=0$
  is determined by the expression $\tau=F(x^2n_{\ze})/(2x^2n_{\ze})$.
 Considering a  special case of  initial condition, which corresponds to the  Planck distribution  $ n_{\ze}(x,0)=n_{\ko}(x,0)=1/(\exp (x)-1)$ with $T_{\rm ph}>T_{\rm e}$, one obtains for
 solution~(\ref{z2}):
\begin{equation}\label{z3}
 n_{\ze}=\frac{1}{2x^2t_1}\left[\left(2t_1-1-\frac{x}{2}\right)+\left(\left( 2t_1-1-\frac{x}{2}\right)^2+4t_1x\right)^\frac{1}{2}\right]
\end{equation}
in the frequency region which corresponds to large occupation number, i.e., for $x<<1$.
\vspace*{1cm}

{\bf Vlasov plasma oscillations}\\ \\
\noindent After this preliminaries we consider plasma consisting of charged  particles with positions ${\r}$ and momenta ${{\p}=m{\v}}$ moving in ${\mathbf{R}}^3$.
The Vlasov equation~\cite{AAVl} for the particle distribution function
(plasma density) $f({\r,\p,}{t})$ is
\begin{equation}\label{a2}
\frac{\partial f}{\partial t}+\frac{{\p}}{m}\cdot \frac{\partial f}{\partial{\r}}-
\frac{\partial(U+\widetilde{U})}{\partial\r}\cdot \frac {\partial f}{\partial{\p}}=0.
\end{equation}
Here $U$ is an external potential and $\widetilde{U}$ is an electrostatic self-consistent plasma potential which is defined by
$\widetilde{U}(t,{\r})=(N_0/V)\int\Phi({\r}-{\r}^\prime) f({\r}^\prime,{\p}^\prime,{t})d^3r^\prime d^3p^\prime$,
 where $\Phi{(\r)}= {e}/{r}$.
\noindent
Let us present the eventual solution of~(\ref{a2}) in the following perturbational form
\begin{equation}\label{a3}
f({\r,\p,}{t})=f_{0}({\p})+\mu f^*({\r,\p,}{t}), \qquad \frac{\mu f^*}{f_0}\ll 1,
\end{equation}
\noindent where $f_0({\p})$ is the space homogeneous Maxwell distribution for electrons.
Then the fluctuating electrostatic field can be written as
\begin{equation}\label{a4}
{\mathbf{E}}({\r},t)= -\frac{\partial}{\partial{\r}} \int \frac{e}{|{\r}-{\r}^\prime|}f^*({{\r}^\prime,{\p}^\prime},t) d^3r^\prime d^3p^\prime .
\end{equation}
From Eq.~(\ref{a4}) we see that ${\mathbf{E}}$ and $f^*$
are of the same order. In what follows, we use the velocities $\v=\p/m$ instead of the momenta. Then using the form of Eq.~(\ref{a3}) we get the linearized Vlasov equation of the order ${O}(\mu)$
\begin{eqnarray}\label{a5}
\frac{\partial f^*}{\partial t}&+&{\v}\cdot \frac{\partial f^*}{\partial{\r}}-\frac{e{\mathbf{E}}}{m}\cdot\frac{\partial f_0}{\partial{\v}}=0,\\
 \mbox{div}{\mathbf{E}}  &=& -4\pi e \int f^*({\r,\v,}{t})d^3v.
\end{eqnarray}
Here
$$
f_0({\v})=\left(\frac{m}{2\pi kT}\right)^{3/2}\exp\left(-\frac{m{\v}^2}{2kT}\right);\quad\frac{\partial f_0}{\partial{\v}}=-\frac{m}{kT}\,{\v}\cdot f_0(\v).\nonumber\\
$$
To be specific consider plasma oscillations propagating
in $\mathrm{r_1}$-direction:
\begin{eqnarray}\label{a6}
f^*({\r,\v,}{t}) &=& f_{k\omega}({\v})\exp(-i\omega t+i k \mathrm{r_1}),\\
E_{\mathrm{r_1}}({\r},t) &=& E_{k\omega} \exp(-i\omega t+i k \mathrm{r_1})\nonumber,\\
E_{\mathrm{r_2}} &=& E_{\mathrm{r_3}}=0.\nonumber
\end{eqnarray}
In particular we are interested the case of the long wave-length limit for $\lambda=2\pi/k$.
Then following Kvasnikov~\cite{Vl},
we get from Eqs.~(\ref{a6}) in this limit $(k\rightarrow 0)$:
\begin{equation}\label{a7}
f^*({\v,}{t}) =f_{0\omega}({\v})\exp(-i\omega t)=\varphi(t)v_1 f_0({\v}), \quad
\varphi(t)=-\frac{e}{\omega_0 kT}\sin\omega_0 t.
\end{equation}
Here $\omega_0$ is Langmuir frequency. For the photon density we look for the representation which is a perturbation of the Kompaneets solution by the  Vlasov oscillations:
\begin{equation}\label{b1}
 n(x,{t_1})=n_{\ko}(x,t_1)+\mu {n}^*(x,{t_1})\quad \mbox{with}\quad {n}^*(x,{0})=0.
\end{equation}
Here $n_{\ko}(x,t_1)$ is the solution of Kompaneets equation~(\ref{kz1}).
 To proceed further we return to Eq.~(\ref{a1}) for the distribution function of photons, but instead of the Maxwell function distribution for the electrons, we consider
the perturbed solution~(\ref{a3}), where $f^*$ is the solution
of Vlasov equation~(\ref{a5}). Then we obtain for unknown function ${n}^*(x,{t_1})$, the
 following linearized equation:
\begin{eqnarray}\label{b2}
 \frac{\partial n^*}{\partial t_1}&\!=\!&\gamma_2\frac{{\partial}^2n^*}{\partial x^2}+\left(\gamma_1+2\gamma_2(n_{\ko}+1)\right)\frac{\partial n^*}{\partial x}\nonumber\\
 &\!+\!&\left(\gamma_1(1+2n_{\ko})\!+\!\gamma_2\left(1+2n_{\ko}+2\frac{\partial n_{\ko}}{\partial x}\right)\right)n^*\!+\!\Phi({v_1},x,t_1),
\end{eqnarray}

where

\begin{eqnarray}\label{korr1}
&\!\!\!&\Phi({v_1},x,t_1)=
\left(\!\frac{\partial n_{\ko}}{\partial x}+n_{\ko}(1+n_{\ko})\right)\frac{\hbar}{kT}\int\! d^3 p\int\! dW \varphi(t)v_1f_0({\v})\Delta\!+\nonumber\\
&\!\!\!\!&\left(\!\frac{{\partial}^2 n_{\ko}}{\partial x^2}\!+\!n_{\ko}(1\!+\!n_{\ko})\!+\!2\frac{\!\partial n_{\ko}}{\partial x}(\!1\!+\!n_{\ko})\!\right)\!
\frac{\hbar^2}{(kT)^2}\int\!\! d^3 p\!\! \int\!\!dW\varphi(t)\!\frac{v_1}{2}f_0({\v})\Delta^2,
\end{eqnarray}

and
\begin{eqnarray*}
\gamma_1&=&\frac{\hbar}{kT}\int d^3 v \int dW f_0({\v})\Delta=x(4-x),\\
\gamma_2&=&\frac{1}{2}\left(\frac{\hbar}{kT}\right)^2\int d^3v \int dW f_0({\v})){\Delta}^2=x^2.
\end{eqnarray*}
Equation~(\ref{b2}) is a linear parabolic equation with non-constant coefficients. Nevertheless, one can find
solutions in some special cases.
Consider them only for some partial case  when the "shift" term ${\partial n^*}/{\partial x}$ and the "force" which proportional to  $n^*$ are much  smaller than the diffusion term. Then
\begin{equation}\label{b6}
 \frac{\partial n^*}{\partial t_1}=x^2\frac{{\partial}^2n^*}{\partial x^2}+\Phi({v_1},x,t_1).
\end{equation}
In the framework of the Zel'dovich - Levich approximation
 $n_{\ko}\gg 1$ and $n_{\ko}^2\gg \left|\partial n_{\ko}/\partial x\right|$, it is straightforward to find that\\
 $\Phi(v_1,x,t_1)=-\varphi(t_1)\, x\, {n_{\ze}^2(x,t_1)}$,
 where $n_{\ze}(x,t_1)$ has the form~(\ref{z3}).
To find solution of Eq.~(\ref{b6}), let us  change  variables:
 $ z=-\ln x$, $\tau=t_1$, and $\widetilde{n^*}(z,\tau)=n^*(x,t_1)$. Then the equation for $\widetilde{n^*}(z,\tau)$ takes the form
\begin{eqnarray}\label{d1}
&& \frac{\partial\widetilde{n^*}}{\partial\tau}=\frac{{\partial}^2\widetilde{n^*}}{\partial{z}^2}+
\frac{\partial\widetilde{n^*}}{\partial z}+\Phi(v_1,z,\tau), \quad -\infty<z<+\infty,\quad 0<\tau<+\infty\nonumber\\
&& \widetilde{n^*}(z,0)=0
\end{eqnarray}
 After the second change of variables: $\widetilde{n^*}(z,\tau)=\exp(-z/2-\tau/4)\nu(z,\tau)$, we get for $\nu(z,\tau)$ diffusion equation:
\begin{equation}\label{d2}
 \frac{\partial \nu}{\partial\tau}= \frac{{\partial}^2 \nu}{\partial z^2} + \Phi(v_1,z,\tau)\exp(z/2+\tau/4),{\hspace{1cm}}\nu(z,0)=0.
\end{equation}
The solution of~Eq.~(\ref{d2}) is
\begin{eqnarray}
 {\widetilde{n}}^*(z,\tau)&=& {\rm e}^{(-\frac{z}{2}-\frac{\tau}{4})}\frac{A}{2\sqrt{\pi}} \int_0^\tau d\tau^\prime\int_{-\infty}^{\infty}dz^\prime \sin\omega_0\tau^\prime n^2_{\ze}(z^\prime,\tau^\prime) \times \nonumber \\
 && \qquad \qquad \qquad \qquad  {\rm e}^{(\frac{3z}{2}+\frac{\tau}{4})}{\rm e}^{\frac{(z-z^\prime)^2}{4(\tau-\tau^\prime)}}(\sqrt{\tau-\tau^\prime})^{-1}\label{d3}
\end{eqnarray}
where $A={el}/(\omega_0 kTc)$. \\ To calculate integral~(\ref{d3}) we consider $n_{\ze}(x,t_1)$ in the following two limit cases ($0<x<1$):
\begin{itemize}
\item The first one is the case of small times $t_1<1/2+x/4$.
Then it is straightforward to show that $4t_1x/(2t_1-1-\frac{x}{2})^2<1$, and
 $n_{\ze}(x,t_1)=\frac{1}{x(1+x/2-2t_1)}$. For simplicity let us consider the Zel'dovoch-Levich approximation, when $n_{\ze}(x,t_1)=\frac{1}{x(1-2t_1)}$. Then one obtains
 $\Phi(v_1,x,t_1)=A\sin\omega_0 t_1\frac{1}{x(1-2t_1)^2}$.\\
\textit{Remark}: In this approximation the last formula for $n_{\ze}$ manifest a singularity at $t_1=1/2$. Zel'dovich and Levich have mentioned that a kind of photon Bose condensation is occurring just at this moment, see also~\cite{YP}, \cite{gruz}. They have found that it is a minimal critical time in which the photon state with $x=0$
starts filling up.\\
For time $t_1<1/2$ one obtains from~(\ref{d3}) the solution of Eq.~(\ref{b6}):
\begin{eqnarray*}
&& n^*(x,t_1)=\frac{A}{2x}(1+2t_1)\int_{1-2t_1}^1\frac{\sin(\frac{\omega_0}{2}-\frac{\omega_0}{2}\eta)}{\eta} d\eta=\nonumber\\
&&=\frac{A}{2x}(1+2t_1)\left[\sin\frac{\omega_0}{2}\int_{\frac{\omega_0}{2}(1-2t_1)}^{\frac{\omega_0}{2}}\frac{\cos\omega_0 \eta^\prime}{\eta^\prime} d\eta^\prime \right.
\\
&& \qquad \qquad \qquad \qquad  \left. -\cos\frac{\omega_0}{2}\int_{\frac{\omega_0}{2}(1-2t_1)}^{\frac{\omega_0}{2}}\frac{\sin\omega_0\eta^\prime}{\eta^\prime} d\eta^\prime\right].
\end{eqnarray*}
\item For the second limit case, $t\gg 1$, one gets  $n_{\ze}(x,t_1)=2/x^2$
with $\Phi(v_1,x,t_1)=4A\sin\omega_0 t_1$. Then
\begin{equation}\label{r2}
   n^*(x,t_1)\!=\!\frac{el}{\omega_0kT} \frac{(12\sin\omega_0 t_1-\omega_0\cos \omega_0t_1)}{x^3(144+\omega_0^2)},\, x=\frac{\hbar\omega}{kT},\, t_1=\frac{mc^2}{kT}\frac{l}{c}t.\nonumber
\end{equation}
\end{itemize}
We notice that, in contrast to the first case, the behavior the photon distribution when $x$ goes to zero is different from the  Planck distribution. It is in fact a dominant contribution as $x$ approaches zero.\\ \\

{\bf Conclusion}\\

\noindent We found the photon function distribution  perturbed by the Compton scattering in plasma is:
\begin{equation}\label{end}
 n(x,{t_1})=n_{\ze}(x,t_1)+\mu {n}^*(x,{t_1}).
\end{equation}
We have shown that the solution exhibited by Zel'dovitch-Levich might be unstable at sufficiently large times around $x=0$ where photon Bose condensation occurs. This instability is generated by the diffusion term in the equation for the perturbation of the photon density by plasma wave oscillations.
\begin{center}
{\bf\small Acknowledgments}
\end{center}
We take the opportunity to thank E.A. Dynin for fruitful discussions.

\end{document}